\documentclass[a4paper,10pt]{article}

\usepackage[utf8]{inputenc}

\usepackage{hyperref}
\usepackage{amssymb}
\usepackage{graphicx}

\newcommand\email[1]{{\tt\href{mailto:#1}{#1}}}

\title{\huge{Acoustic properties of glacial ice for neutrino detection and the Enceladus Explorer}}

\author{\Large{Klaus Helbing$^a$, Ruth Hoffmann$^a$\footnote{Speaker} , Uwe Naumann$^a$,}\\
  \Large{Dmitry Eliseev$^b$, Dirk Heinen$^b$, Franziska Scholz$^b$,}\\
  \Large{Christopher Wiebusch$^b$ and Simon Zierke$^b$}\\\\
  \small{\llap{$^a$} Dept. of Physics, University of Wuppertal, 42119 Wuppertal, Germany}\\
  \small{\llap{$^b$}III. Physikalisches Institut, RWTH Aachen University, 52056 Aachen, Germany}\\
  E-mail: \email{rhoffman@uni-wuppertal.de}}
  
\date{}

\begin{document}

\maketitle
\thispagestyle{empty}

Ultra high energy neutrinos may be observed in ice by the emission of acoustic signals. The SPATS detector has investigated the possibility of observing GZK-neutrinos in the clear ice near the South Pole at the IceCube detector site. To explore other potential detection sites glacial ice in the Alps and in Antarctica has been surveyed for its acoustical properties. \\
The purpose of the Enceladus Explorer (EnEx), on the other hand, is the search for extraterrestrial life on the Saturn moon Enceladus. Here acoustics is used to maneuver a subsurface probe inside the ice by trilateration of signals. A system of acoustic transducers has been developed to study both applications. \\
In the south polar region of the moon Enceladus there are secluded crevasses. These are filled with liquid water, probably heated by tidal forces due to the short distance to Saturn. We intend to take a sample of water from these crevasses by using a combination of a melt down and steering probe called IceMole (IM). Maneuvering IM requires a good understanding of ice properties such as the speed of sound, the attenuation of acoustic signals in ice, their directional dependencies and their dependence on different frequencies. The technology developed for this positioning system could also contribute to the design of future large scale acoustic neutrino detectors. \\
We present our analysis methods and the findings on attenuation, sound speed, and frequency response obtained at several sites in the Alps and Antarctica.
\\ \\ \\ \\ \\ \\
\small{The 34th International Cosmic Ray Conference,}\\
\small{30 July- 6 August, 2015}\\
\small{The Hague, The Netherlands}

\newpage
\section{Introduction}

\subsection{Ultra high energy neutrinos and acoustic signals}

In order to observe neutrinos with ultra high energies ($E \gtrsim 10^{18}\,\textrm{eV}$) the instrumentation of detector volumes of at least several tens of cubic kilometers is necessary. The thermo acoustic effect is %(Figure \ref{ThermoAcousticEffect_pic}) 
one detection option \cite{AcousticDetection, SPATS, AND_Background, SPATSattenuation}. The SPATS experiment demonstrated the feasibility of an acoustic neutrino detector as an extension of IceCube in the clear ice at the South Pole where the attenuation of acoustic signals is in the range of a several hundred meters. In this work we investigated the possibility to build such a detector in the ice of normal glaciers as found in the Alps or at the outer parts of the Antarctic continent.

\subsection{The Enceladus Explorer and acoustic positioning}

\begin{figure}[h]
  \centering
  \includegraphics[width=0.6\textwidth]{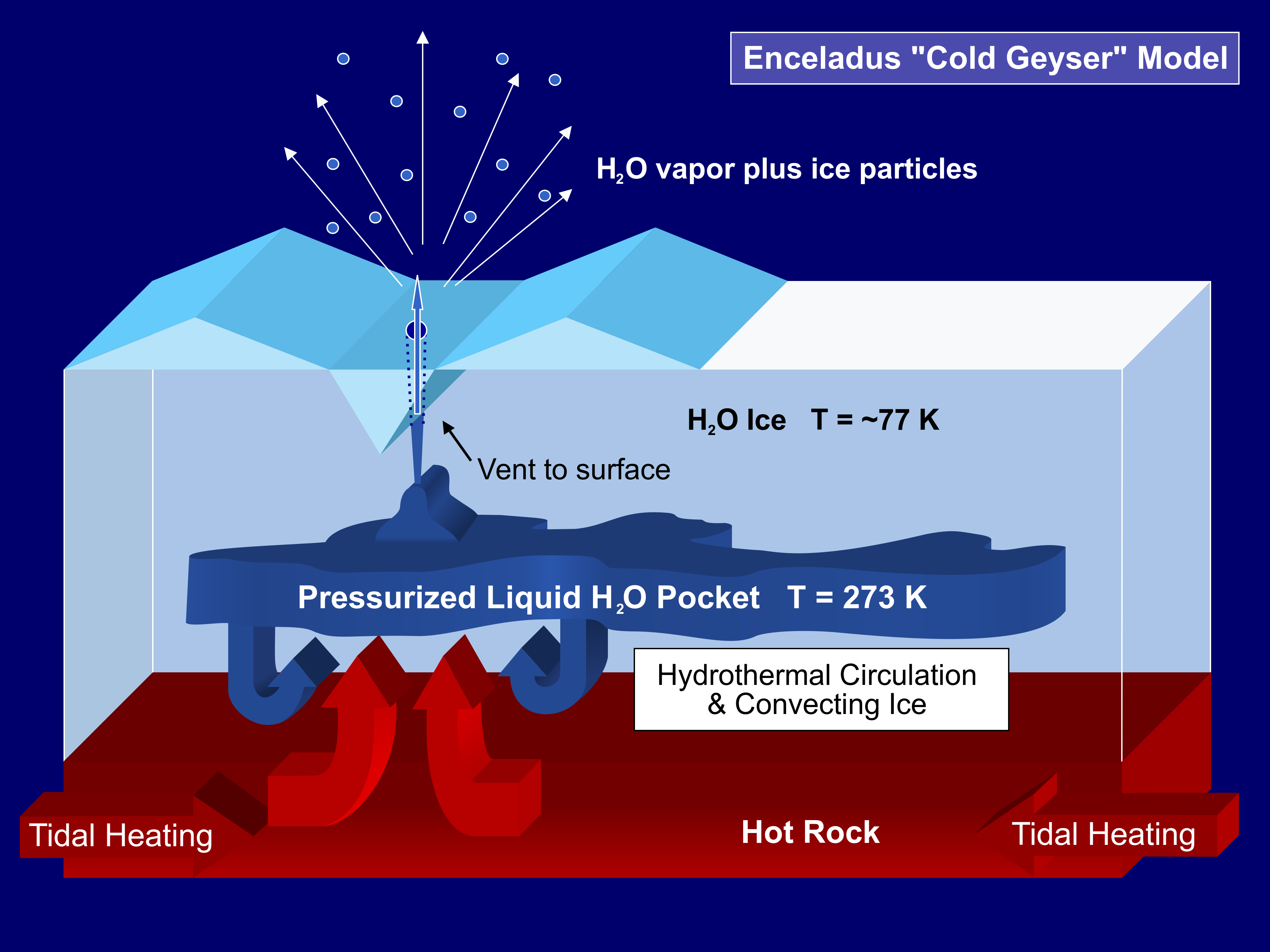}
  \caption{Cold Geyser model on the Saturn moon Enceladus \cite{ColdGeyser}.}
  \label{Cold_Geyser_Model_pic}
\end{figure}

Cassini discovered "Tiger Stripes" in the south polar region of the Saturn moon Enceladus. These are, compared to the surrounding ice, hot crevasses through which liquid water leaks and evaporates at the surface to feed the E-ring of Saturn. The occurrence of liquid water in combination with energy from tidal forces within the moon (see Figure \ref{Cold_Geyser_Model_pic}) makes Enceladus a good candidate for the search for extraterrestrial life. 

\begin{figure}[ht]
  \centering
  \includegraphics[height=0.29\textwidth]{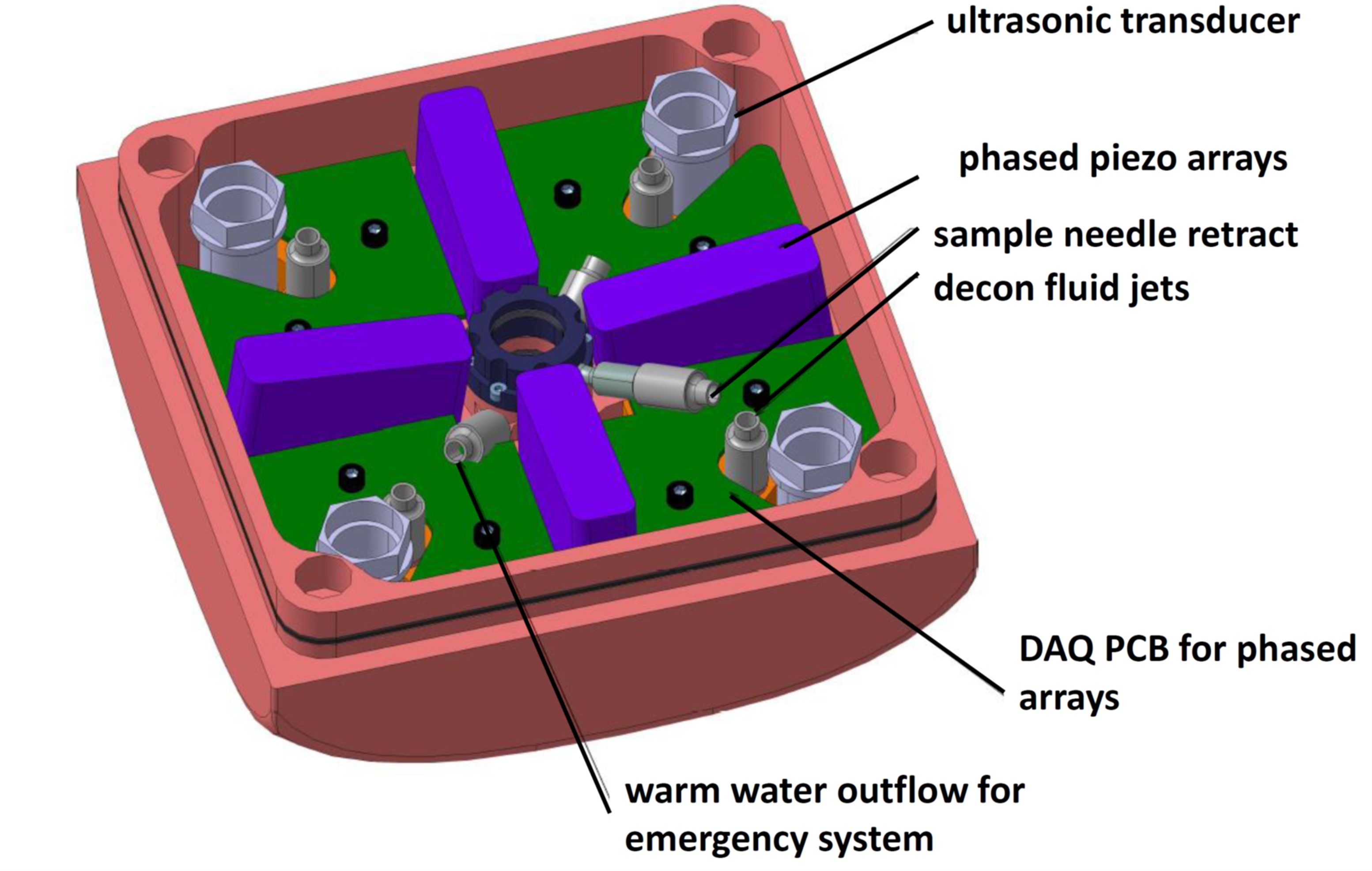}
  \includegraphics[height=0.29\textwidth]{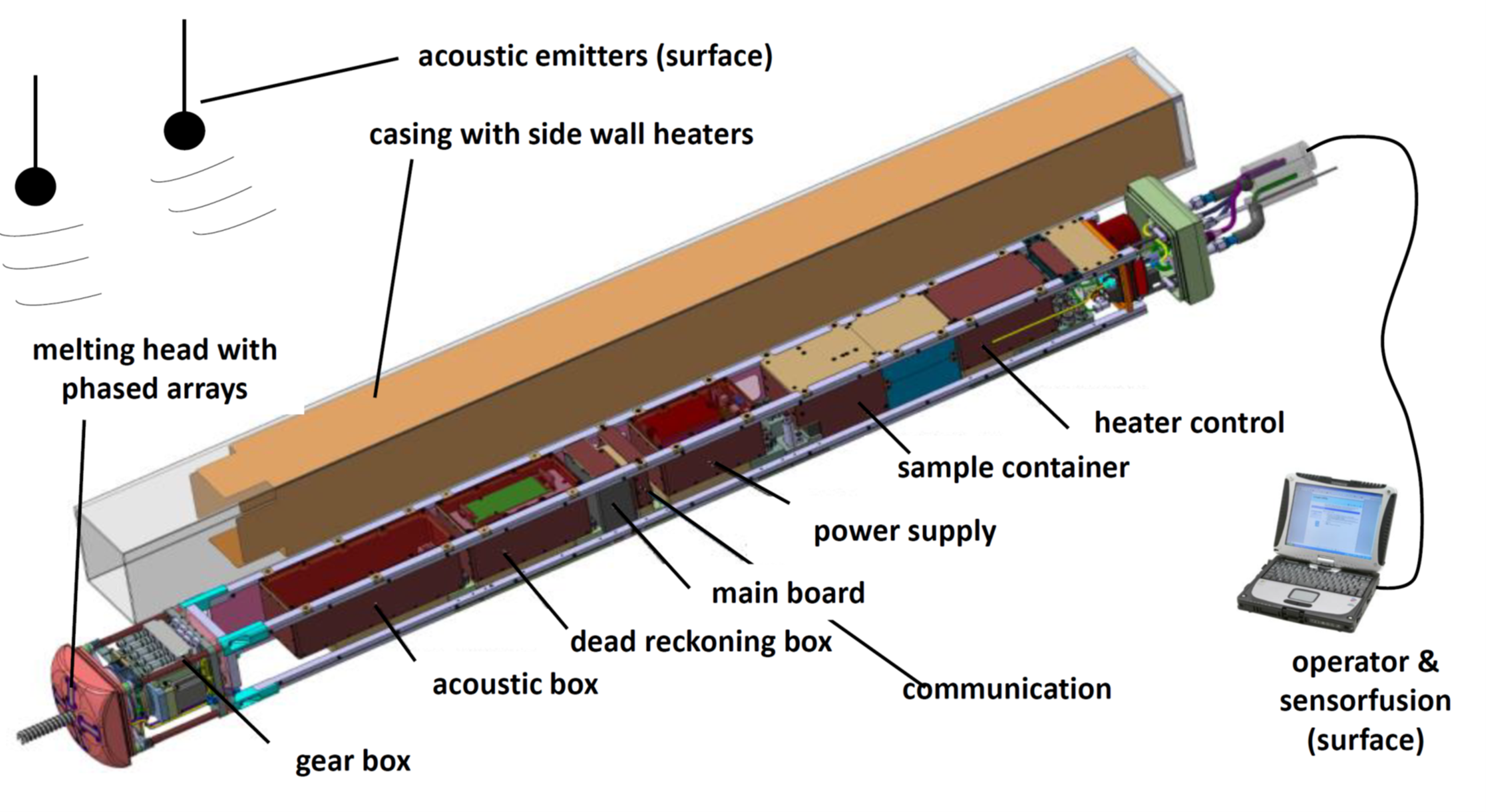}
  \caption{Overview drawing of the EnEx-IceMole (right) with all subsystems and a view of the inside of the melting head with the acoustic receivers (left) \cite{IceMole}.}
  \label{IceMole_pic}
\end{figure}

The idea behind the Enceladus Explorer is to use a steerable melt down probe such as the IceMole shown in Figure \ref{IceMole_pic} and take a water sample directly from one of the crevasses. The best setting for terrestrial field tests can be found at the Blood Falls in the Dry Valleys of Antarctica where this whole concept was tested. To achieve this goal a positioning system is needed and one idea is to use acoustic pulses and trilateration algorithms to determine the position of the probe. 

\newpage
\section{Acoustic positioning system (APS)}

\begin{figure}[h]
  \centering
  \includegraphics[width=0.7\textwidth]{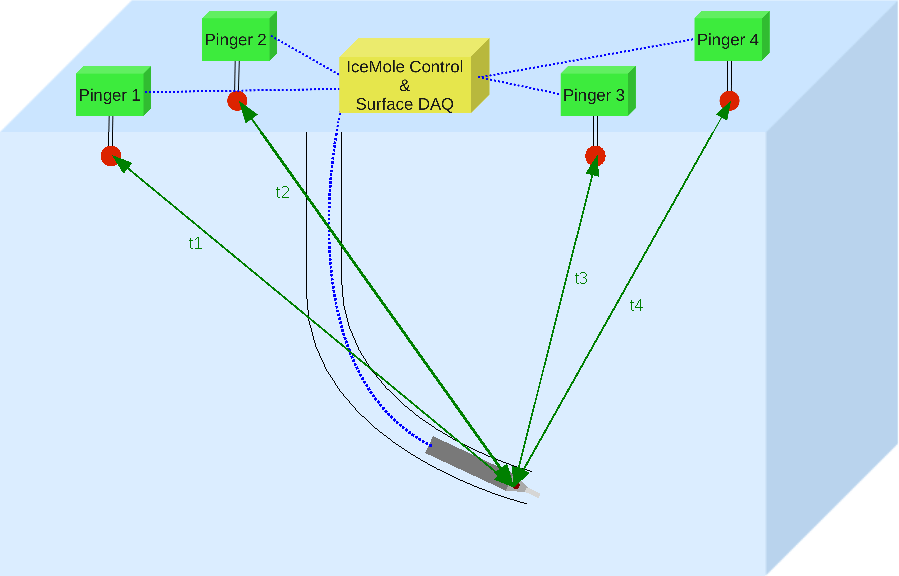}
  \caption{Principle of the acoustic positioning system.}
  \label{APS_Prinzip_pic}
\end{figure}

The concept of the APS is shown in Figure \ref{APS_Prinzip_pic}. On the surface of the glacier the transducers are located at known positions from where they send the acoustic pulses sequentially to the IceMole. By measuring the signal propagation times, and assuming a known speed of sound, the distances between each transducer and the IceMole can be calculated and its absolute position then derived from these distances. 

The APS for EnEx consists of six acoustic transducers, each connected via front-end electronics that are used for amplification of sent and received signals to a central transducer unit. It also generates the control signals and synchronizes with the receivers in the IceMole head. For the transducers a hollow PZT (Lead zirconate titanate) sphere with a diameter of $10\,\textrm{cm}$ is used with a resonance frequency of $18\,\textrm{kHz}$. The speed of sound in amorphic glacial ice is neither exactly known nor constant which is why additional measurements for the determination of the ice properties became necessary. Signals between all pairs of transducers have been taken and a complementary pair of transducers were used for depth dependent measurements.

\section{Field test results}

In order to increase our knowledge about sound propagation in glacial ice several field tests were performed over the last three years both in temperate and cold ice. Measurements in Switzerland (Morteratsch and Pers glacier) were focused on the coupling between transducers and ice, directional dependencies and frequency and range optimization. The goal of the tests in Antarctica (Canada and Taylor glacier) was to study the influence of lower temperatures besides the actual positioning of the IceMole.

\subsection{Speed of sound}

To derive the signal propagation times from the recorded waveforms different approaches have been tested to find a stable method that works automated for all recorded data. First, multiple pulses in the same configuration have been averaged. Then the running mean is computed over a half wave of the Hilbert envelope of the averaged signals and then a threshold at $20\,\textrm{\%}$ of the maximum in the whole time window is applied.

\begin{figure}[h]
  \centering
  \includegraphics[width=0.45\textwidth, height=0.45\textwidth]{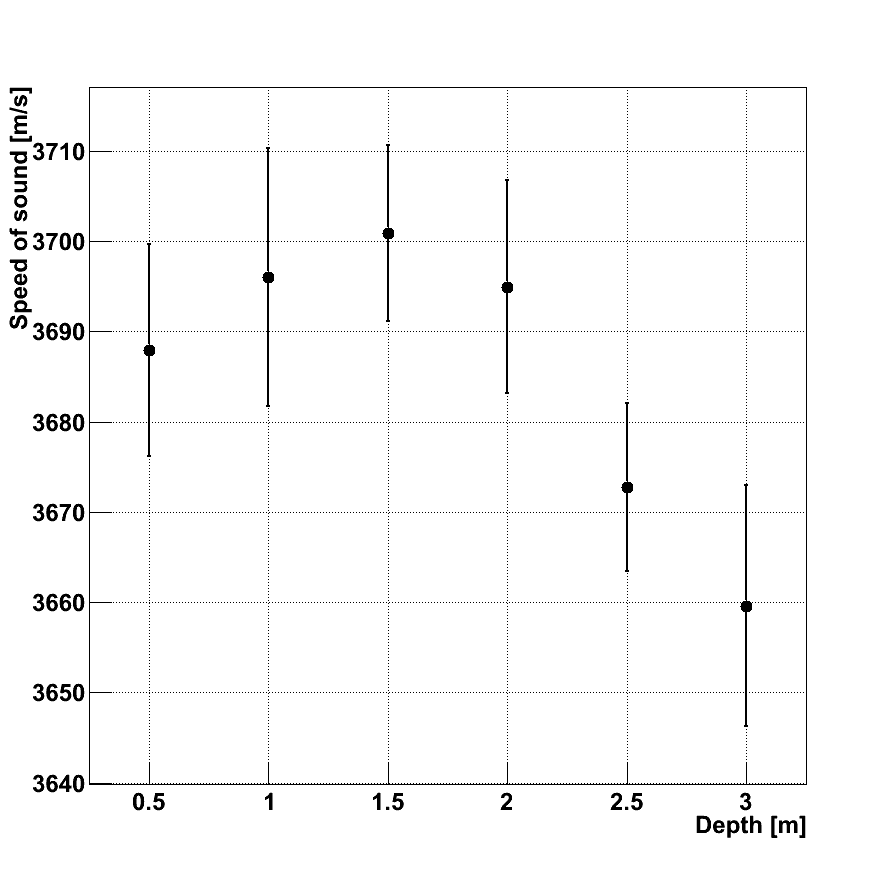}
  \includegraphics[width=0.45\textwidth, height=0.45\textwidth]{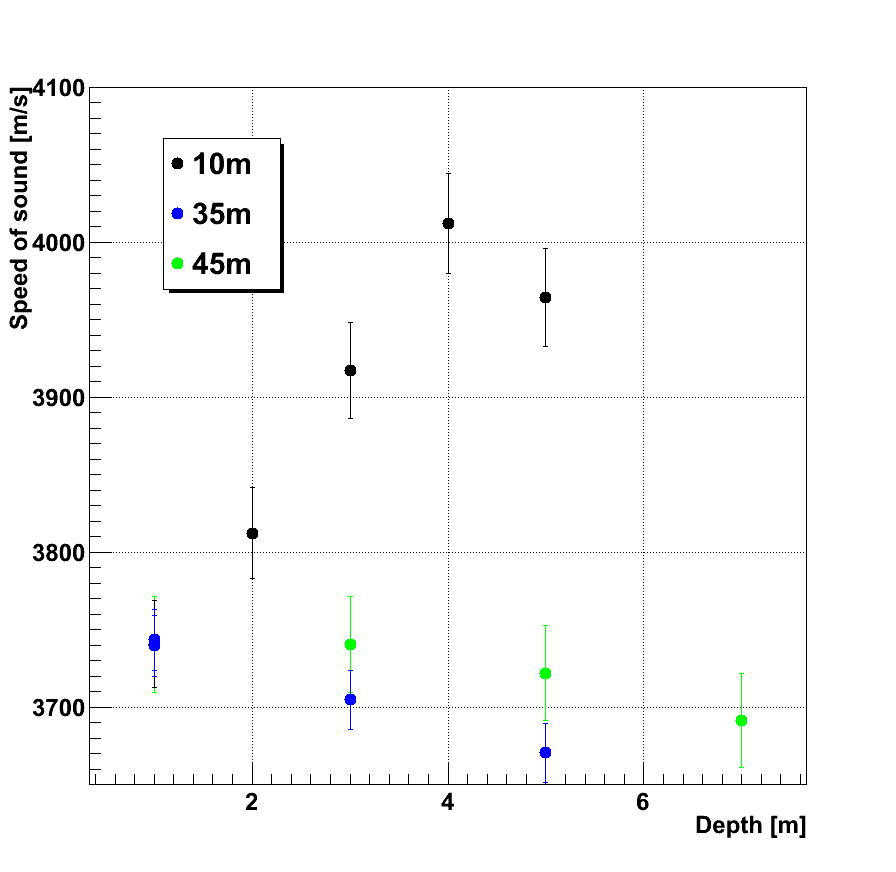}
  \caption{Depth measurements of the speed of sound from two field tests on the Morteratsch glacier. Each point was measured by inserting two transducers in different water filled deep holes at the same depth. The data on the right side shows a differing behavior at larger distances. The profile on the left was measured at a distance of $11\,\textrm{m}$ and shows a comparable behavior like the profile at $10\,\textrm{m}$ on the right. The data on the left was recorded with a frequency of $30\,\textrm{kHz}$ and the data on the right side at $18\,\textrm{kHz}$.}
  \label{DepthProfiles_pic}
\end{figure}

During most field tests the positions of the transducers were determined by differential GPS with a precision better than $2\,\textrm{cm}$. Only during the first test campaigns on the Morteratsch glacier and on the Pers glacier the positions were measured using a compass, a protractor, and measuring tape which in some cases over large distances ($\sim100\,\textrm{m}$) lead to uncertainties up to $1\,\textrm{m}$ on the position on the ice. The depth in the ice was determined with uncertainties up to $5\,\textrm{cm}$. The default frequency for most measurements was $18\,\textrm{kHz}$ which corresponds to the resonance frequency of the transducers.

\begin{figure}[ht]
  \centering
  \includegraphics[width=0.45\textwidth, height=0.45\textwidth]{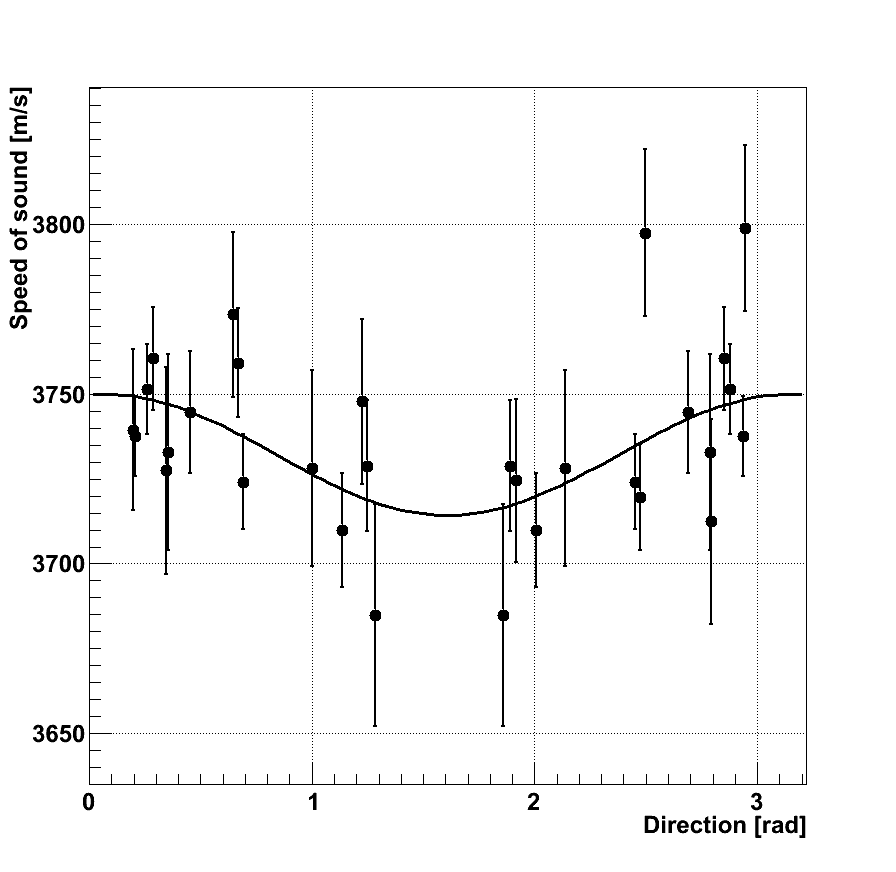}
  \includegraphics[width=0.45\textwidth, height=0.45\textwidth]{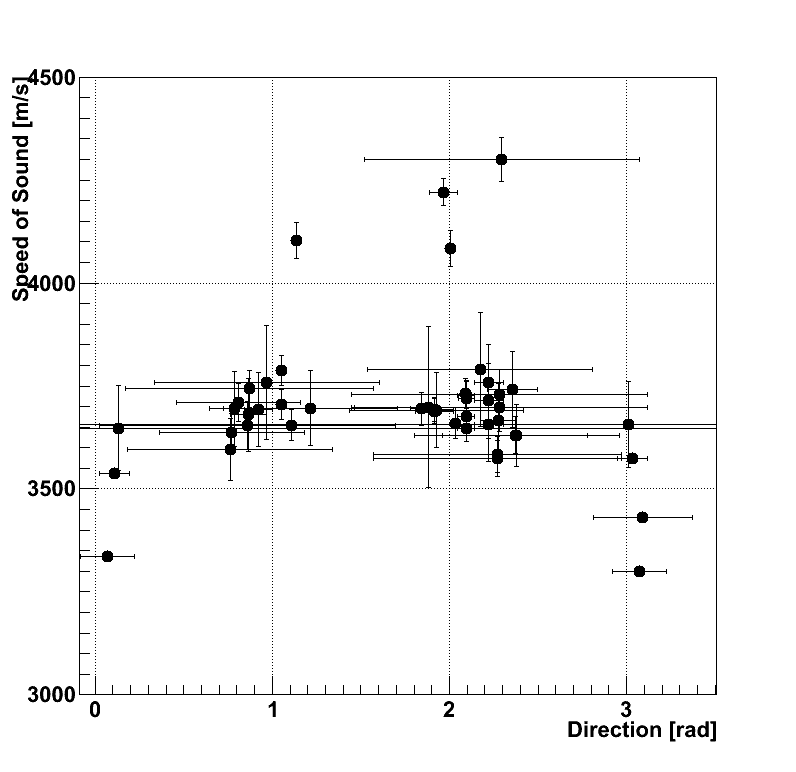}
  \caption{Measurements of the speed of sound in different directions parallel to the surface at a depth of about $1\,\textrm{m}$ from different test sites on temperate glaciers such as the Morteratsch glacier (left) and the Pers glacier (right). The data from the Morteratsch glacier was measured within the array of the APS. The signals on the Pers glacier were measured by using the same pair of transducers in different holes distributed over an area of $7500\,\textrm{m}^2$.}
  \label{DirectionWarm_pic}
\end{figure}

As shown in Figure \ref{DepthProfiles_pic} variations of the speed of sound of up to $10\,\textrm{\%}$ have been observed for a s pair of holes. Varying sound speeds are also observed comparing hole pairs at different distances on the same test site. 

\begin{figure}[h]
  \centering
  \includegraphics[width=0.45\textwidth, height=0.45\textwidth]{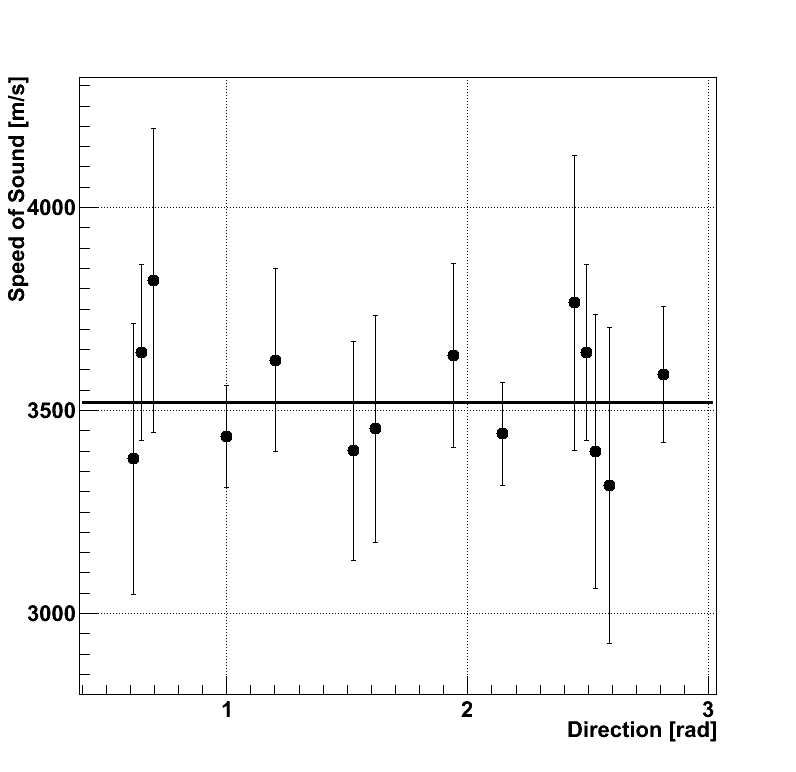}
  \includegraphics[width=0.45\textwidth, height=0.45\textwidth]{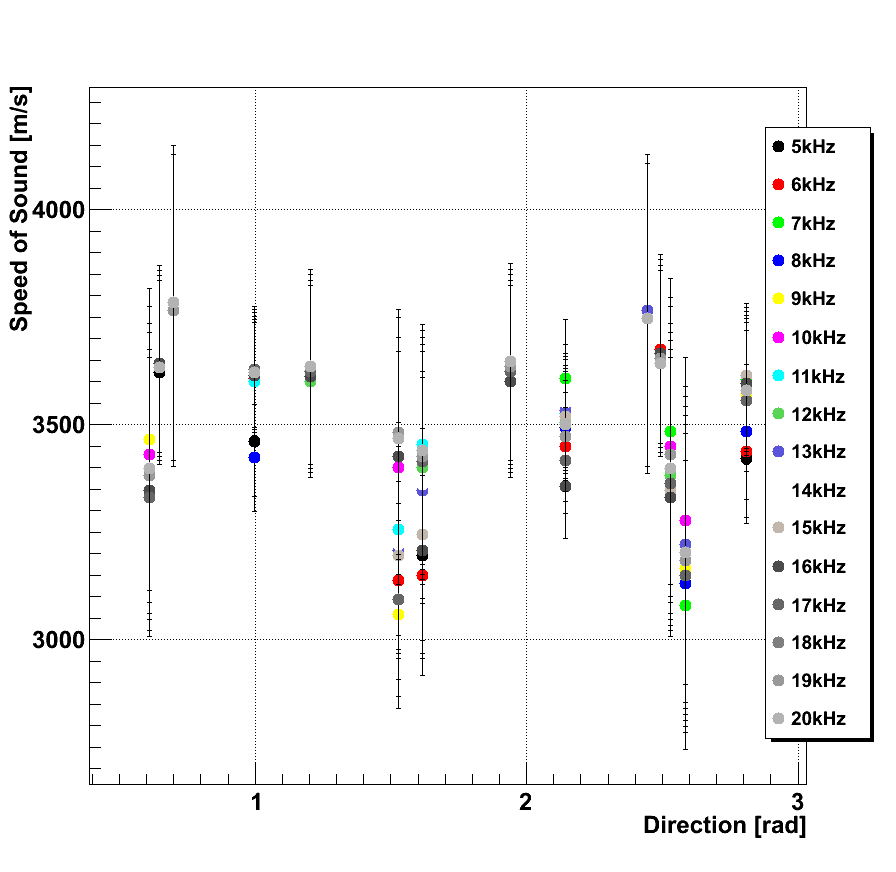}
  \caption{Directional measurements of the speed of sound in cold ice from Blood Falls. The data shown on the left side is comparable to the plots above and the data on the right side shows the same plot for different frequencies. Both plots are derived from data taken within the array.}
  \label{DirectionCold_pic}
\end{figure}

Systematic directional variations parallel to the surface have only been observed on one test site on the Morteratsch glacier (Figure \ref{DirectionWarm_pic}) and are in the order of $5\,\textrm{\%}$. The data from all other test sites is consistent with a speed of sound not depending on the direction as fitted to the data shown in Figure \ref{DirectionCold_pic}. The outliers on the Pers glacier can be explained by the presence of stones (faster) and snow (slower) next to the holes. \newline
The average speed of sound in the cold ice of the Antarctic field tests was systematically lower than the speed of sound in the temperate ice of the Swiss glaciers. On the right side of Figure \ref{DirectionCold_pic} the speed of sound is plotted for frequencies from $5-20\,\textrm{kHz}$ in the same configuration as before. No significant deviations of the speed from that at the resonance frequency were observed.

\subsection{Attenuation}

\begin{figure}[h]
  \centering
  \includegraphics[width=0.45\textwidth, height=0.45\textwidth]{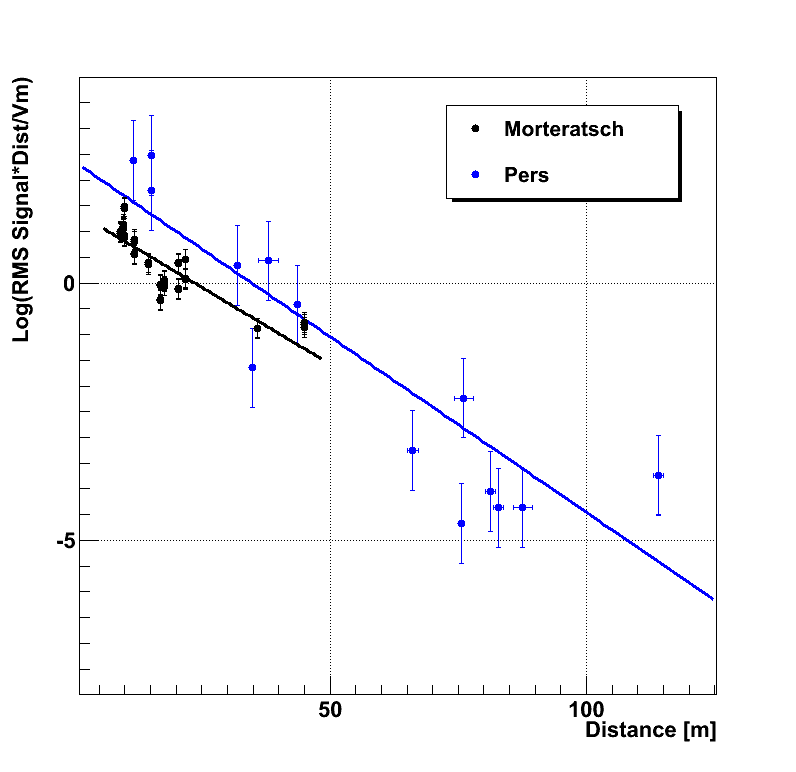}
  \includegraphics[width=0.45\textwidth, height=0.45\textwidth]{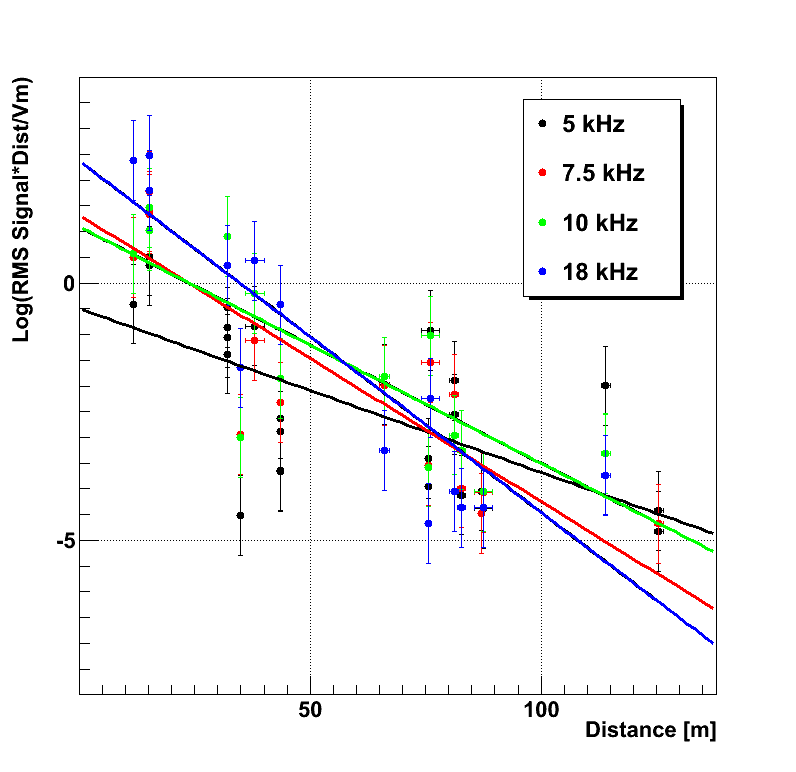}
  \caption{Left: Comparison of the attenuation on the Morteratsch glacier and on the Pers glacier. The plot for the Morteratsch glacier shows the combined measurements of the positioning array and the additional points taken outside of the array. The data from the Pers glacier originates only from the complementary system. Right: Attenuation for different frequencies on the Pers glacier.}
  \label{Attenuation_pic}
\end{figure}

For the determination of the attenuation coefficient it is important to have pairs of transducers that are coupled to the ice in a well defined way, so that the losses due to transition from transducer to ice are comparable for each pair. During the test campaigns in Switzerland the coupling was realized through water filled holes which ensures good reproducibility. During the tests in Antarctica the holes with the transducers refroze which resulted in very different coupling for each transducer, complicating reliable attenuation measurements. Therefore the attenuation of the acoustic signals has only been determined during the second field test on the Morteratsch glacier and on the Pers glacier. \newline
For the estimation of the uncertainties of the sensor coupling the variance of signal widths at comparable separations was used to calculate relative uncertainties. Since the coupling is independent of the distance, these relative uncertainties were then used to calculate the absolute uncertainties for all data points at all distances and then, combined with the uncertainties on the distance, transferred to the errorbars shown in Figure \ref{Attenuation_pic}. 

The computation of the attenuation coefficients is based on the estimation of the square root of the energy content of the received signals. In order to get a reliable measure for this, one can use different variables such as the maximal amplitude, the area beneath the curve, or the width of the signal, defined by the RMS. In our case the width of the signal appears to be the most stable. For the calculation only the first 32 waves are considered which corresponds to the duration of the sent pulses. The Logarithm of the RMS of this region multiplied by the distance for correction of the losses due to spherical propagation of the pulses is shown in Figure \ref{Attenuation_pic}. The attenuation coefficients and lengths calculated from the fits are summarized in Table \ref{Attenuation_tab}. 

\begin{table}[htbp]
\caption{Attenuation lengths from the fits.}
\begin{center}
\begin{tabular}{|c|c|c|c|}
\hline
f [kHz] & k [1/m] & $\lambda$ [m] & $\chi^2/NDF$ \\ \hline\hline
\multicolumn{4}{|c|}{Pers glacier} \\ \hline
5.0 & $0.0318 \pm 0.0047$ & $31.4 \pm 4.7$ & 2.75 \\ \hline
7.5 & $0.0557 \pm 0.0059$ & $18.0 \pm 1.9$ & 2.12 \\ \hline
10.0 & $0.0459 \pm 0.0060$ & $21.8 \pm 2.8$ & 2.17 \\ \hline
18.0 & $0.0683 \pm 0.0060$ & $14.6 \pm 1.3$ & 2.40 \\ \hline
\multicolumn{4}{|c|}{Morteratsch glacier} \\ \hline
18.0 & $0.0596 \pm 0.0027$ & $16.8 \pm 0.8$ & 3.30 \\ \hline
\end{tabular}
\end{center}
\label{Attenuation_tab}
\end{table}

The Morteratsch glacier shows slightly but not significant smaller attenuation at $18\,\textrm{kHz}$ compared to the Pers glacier. An important difference between these glaciers was that the ice on the Pers glacier was still covered with $1-2\,\textrm{m}$ snow during the test campaign. This may lead to colder and therefore dryer ice. On the other hand the test site on the Morteratsch was very close to a large crevasse in a region where the ice could be more fissured than on the Pers glacier. However, a final conclusion would require a more detailed investigation. 

The comparison of the attenuation for different frequencies (Figure \ref{Attenuation_pic} right) shows that there is a clear dependency. The attenuation length for $18\,\textrm{kHz}$ is significantly smaller than for lower frequencies.

\section{Summary}

All reliable measured attenuation coefficients are above $0.03\,\textrm{m}^{-1}$ which corresponds to an upper limit of the attenuation length of $\sim30\,\textrm{m}$. This indicates strong attenuation losses in fast flowing, cirque glacial ice which would make the detection of ultra-high energy neutrinos very challenging. However using frequency matching between APS and ice leads to a sufficient range for the positioning of a melting probe. The observed variations of the speed of sound are below $10\,\textrm{\%}$ but need to be measured on every field of operation to optimize the precision of the positioning.

\end{document}